\documentclass[12pt,a4paper]{article}
\usepackage{amsmath}
\usepackage{amssymb}
\usepackage{amsthm}
\usepackage{float}
\usepackage{amsfonts}
\usepackage{subfig}
\usepackage{graphicx}
\usepackage{verbatim}
\usepackage[left=2cm,right=2cm,top=3cm,bottom=2.5cm]{geometry}
\usepackage[numbers]{natbib}
\usepackage{notoccite}
\usepackage[utf8]{inputenc}
\usepackage[usenames,dvipsnames,svgnames]{xcolor}
\usepackage[colorlinks=true,
      linkcolor=red,
      urlcolor=gray,
      citecolor=blue]{hyperref}

\def\myalign#1{%
  \def\trule{\noalign{\smallskip\hrule\medskip}}
  \def\nebc{\nearrow\bigcup}
  \def\sebc{\searrow\bigcup}
  \def\pminf{{}_{-\infty}|^{+\infty}}
  \let\Inf\infty
  \def\amp{&} 
  \vbox{\mathsurround0pt\openup1\jot
    \halign{%
      &$\displaystyle##\hfil\tabskip0pt$&\amp##\tabskip1em\crcr
      \noalign{\hrule height1pt\smallskip}#1\noalign{\smallskip\hrule height1pt}\crcr}}}

\def\ber {\begin{eqnarray}}
\def\eer {\end{eqnarray}}
\newcommand{\be}{\begin{equation}}
\newcommand{\ee}{\end{equation}}

\begin{document}

\begin{center}
 \textbf{Inflationary $f(R)$ cosmologies}
\end{center}
\hfill\newline
Heba Sami$^{1,2}$, Joseph Ntahompagaze$^{2,3,4}$ and Amare Abebe$^{1,2}$\\
Email: hebasami.abdulrahman@gmail.com\\
\hfill\newline
$^{1}$ Center for Space Research, North-West University, South Africa\\
$^{2}$ Department of Physics, North-West University, South Africa\\
$^{3}$ Astronomy and Astrophysics Division, Entoto Observatory and Research Center, Ethiopia\\
$^{4}$ Department of Physics, College of Science and Technology, University of Rwanda, Rwanda \\

\begin{center}
Abstract 
\end{center}
  
 This paper discusses a simple procedure to reconstruct $f(R)$-gravity models from exact 
 cosmological solutions of the Einstein field equations with a non-interacting classical scalar 
 field-and-radiation background. From the kind of inflationary scenario we want, we show how the 
 potential functions can be obtained. We then show how an $f(R)$ gravitational Lagrangian density that 
 mimics the same cosmological expansion as the scalar field-driven inflation of General Relativity can be reconstructed. 
 As a demonstration, we calculate  the slow-roll parameters (the spectral index $n_{s}$ and the tensor-to-scalar ratio $r$)  
 and compare them to the Planck data.\\ 

$keywords:$ $f(R)$ gravity; scalar field, inflation
$PACS:$04.50.Kd, 04.25.Nx, 98.80.-k, 95.36.+x, 98.80.Cq

\section{Introduction}
Cosmological inflation is an early-stage accelerated expansion of the universe, first introduced to solve the horizon and flatness problems \cite{guth1981inflationary}.
The usual approach is to assume that,  in the very early universe,  a scalar field dominated standard matter fields that source the action in General Relativity (GR).  One can claim that the dominance of the scalar 
field in the early universe implies the contribution of the curvature
through the extra degree of freedom that is hidden in the $f(R)$-gravity theories compared to GR-based cosmology.
The study of cosmological inflation in modified gravity such as $f(R)$ theory was pioneered by Starobinsky \cite{starobinsky1980new}, where it was shown that $f(R)$ corrections to the standard GR action can lead to an early phase of de Sitter expansion, and several studies have been conducted since then \cite{nojiri03,nojiri07,nojiri2008future,nojiri2011unified,bamba14,amin2016viable}.
The reconstruction techniques of $f(R)$ Lagrangians from the scalar field are done in different ways
\cite{li2007cosmology,chakraborty2016solving,saez2009modified,nojiri2009cosmological,ntahompagaze2017f,sami2017reconstructing,fara07}.
One can explore how the scalar field that dominates in the inflation epoch relates with the geometry through
the derivation of the gravitational Lagrangian which is constructed from both radiation and scalar field inputs.
Thus the idea of combining the two theories results in the Lagrangians which are purely geometric, hence the curvature 
characteristics during the inflation epoch will be revealed.

In inflation theory there are several types of potentials that have different behaviors \cite{guth1993inflation}. 
The nature of the potential dependence on the scalar field shows how slow-roll situation 
affects the scalar field \cite{linde2015inflationary}. 
In principle for a given potential, one can obtain the expressions for the tensor-to-scalar ratio $r$ and spectral index $n_{s}$. 
These parameters can be determined and  compared to the available observations \cite{linde2015inflationary,slow3,slow4} .
In this work, we point out that for a given Lagrangian, one could actually have  the potential with values of parameters that can be constrained by
$r$ and $n_{s}$  values from observational data.

 The exact scalar field can be used to explore some inflation solutions. In \cite{ellis1991exact}, exact potentials
 for different expansion models were obtained  where the radiation contribution is neglected for de Sitter spacetimes, but it was indicated that one can, in principle, generalize the study to include radiation. In this paper, we do include the radiation contribution and we are only 
limited to two scale factor expansion models namely, exponential and linear. The potentials that correspond to the expansion models 
under consideration are used to obtain the parameters like $r$ and $n_{s}$ after the calculations of $f(R)$ Lagrangians.
One could see how the inclusion of radiation contribution makes the calculations complicated.
The comparison with the Planck Survey results is made where the ranges of the parameters are taken into account.

This paper is organized as follows. In the next section we review the main equations involved in the calculations. In
Section \eqref{EXPONENTIALEX}, we consider the exponential expansion law and we obtain the $f(R)$ Lagrangians. 
In Section \eqref{LINEAREX}, the linear expansion law is taken into consideration and also the Lagrangians are obtained as well. Section \eqref{INFLATIONSEC}
is about slow-roll approximations.
Section \eqref{CONCLUSION} is devoted for discussions and conclusions.

\section{Matter description}
We consider a Friedmann-Lema\^itre-Robertson-Walker (FLRW) background filled with a non-interacting combination  of a classical scalar field and radiation such that the energy-momentum tensor is given in terms of the total energy density $\mu$ and isotropic pressure $p$ as \cite{ellis1991exact}
\begin{equation}
T_{ab}=(\mu+p)u_{a}u_{b}+pg_{ab} \; , 
\end{equation}
where
\ber
&&\mu=\mu_{m}+\mu_{\phi} \; ,\\
&&p=p_{m}+p_{\phi}\; ,
\eer
$g_{ab}$ is the metric tensor and $u_{a}$ is the 4-velocity  vector field of fundamental observers.
The energy density and pressure  of the scalar field are given as \citep{ ellis1991exact, gorini2005chaplygin} 
\ber
&&\mu_{\phi}=\frac{1}{2}\dot{\phi}^{2}+V(\phi)\; , \\
&&p_{\phi}=\frac{1}{2}\dot{\phi}^{2}-V(\phi)\;,
\eer
whereas for radiation, 
$ \mu_r=\frac{M}{a^{4}}\;, $ and $p_r=1/3\mu_r$, $a=a(t)$ being the cosmological scale factor and $M$, a constant of time. Thus, for the total fluid of the cosmic medium, one has 
\ber
&&\mu=\mu_{r}+\frac{1}{2}\dot{\phi}^{2}+V(\phi)\; , \\
&&p=\frac{\mu_{r}}{3}+\frac{1}{2}\dot{\phi}^{2}-V(\phi)\; .  
\eer
The scalar field obeys the Klein-Gordon equation \cite{ellis1991exact} given as
\begin{equation}\label{kge}
\ddot{\phi}+3H\dot{\phi}+\frac{\partial V}{\partial \phi}\; .
\end{equation}
The field equations and conservation equations of the background spacetime are given as
\ber\label{friedmann}
&&3H^{2}+3K=\frac{1}{2}\dot{\phi}^{2}+V(\phi)+\mu_{r} \; ,\\
&&\label{raychadhuri}3\dot{H}+3H^{2}=V(\phi)-\dot{\phi}^{2}-\mu_{r}\; , \\
&& \label{conservation}\dot{\mu}+3H(\mu+p)=0\; , 
\eer
where 
$K\equiv\frac{k}{a^{2}}\;, k=\pm1, 0$ is the spatial curvature
 and 
$ H\equiv\frac{\dot{a}}{a}\; $ is the Hubble (expansion) parameter.
 For non-interacting fluids, we can split Eq. \eqref{conservation} and rewrite
\ber
\dot{\mu}_{r}+4H\mu_{r}=0\;,\\
\dot{\mu}_{\phi}+3H\dot{\phi}^{2}=0\;. 
\eer
We combine Eqs. \eqref{friedmann} and \eqref{raychadhuri} to solve for the potential and the scalar field as 
\begin{equation}\label{potential}
V(\phi)=3H^{2}+2K+\dot{H}-\frac{\mu_{r}}{3} \; , 
\end{equation}
and 
\begin{equation}\label{kineticterm}
\dot{\phi}^{2}=2K-2\dot{H}-\frac{4}{3}\mu_{r}\; . 
\end{equation}
The above two equations \eqref{potential} and \eqref{kineticterm} can be solved once one has the expressions for $H$ and $\dot{H}$ 
with the specification of geometry of the spacetime, $k$.  In the following two sections, we consider two different 
expansion laws and obtain the solutions of the scalar field $\phi$ and the corresponding potential. The reconstruction of $f(R)$ Lagrangian densities will need the definition of the Ricci scalar given as
\begin{equation}\label{Ricciscalar}
R=6(\dot{H}+2H^{2}+K)\; . 
\end{equation}
The action for $f(R)$ gravity theories is given as
\begin{equation}
 \mathcal{A}= \frac{1}{2\kappa} \int d^{4}x\sqrt{-g}\left[ f(R)+2\mathcal{L}_{m}\right]\; ,
\end{equation}
where $\kappa= 8\pi G$ (set to unity from here onwards) and $\mathcal{L}_{m}$ is the matter Lagrangian. The field equations derived from the above action, applying the variational principle with respect to the metric $g_{ab}$, describe the same cosmological dynamics as the Brans-Dicke sub-class of the broader scalar-tensor theories. There are different ways of reconstructing
the $f(R)$ Lagrangians from scalar-tensor theory with the scalar field defined in a different way \cite{li2007cosmology,chakraborty2016solving,amin2016viable,saez2009modified,nojiri2009cosmological}.
Here we define the  Brans-Dicke scalar field $\phi$ as \cite{frolov2008singularity}
\begin{eqnarray}\label{definitionofphi}
\phi=f'-1 \;,
\end{eqnarray}
where $f'=\frac{df}{dR}$,  such that for GR the extra degree of freedom automatically vanishes, {\it i.e.}, $\phi=0$. We can see from this definition that
 \begin{equation}\label{defintionf(R)}
 f(R)= \int (\phi+1)dR+C\;.
\end{equation}

For a specified scale factor $a(t)$, one would solve Eq. \eqref{kineticterm} to get the momentum of the scalar field $\dot{\phi}$.
Its integration with  respect to time gives the expression of the scalar field. To connect the scalar field $\phi$ with the Ricci-scalar $R$,
we have to get $t(R)$ from Eq. \eqref{Ricciscalar}. Then after establishing $\phi(R)$, we can use the definition defined in 
Eq. \eqref{definitionofphi} to obtain the Lagrangian $f(R)$. One can get the potential $V(t)$ from Eq. \eqref{potential} with 
the specification of the scale factor and we first get $t(\phi)$ then we get $V(\phi)$.
In the following, we consider two expansion models namely, exponential and linear models. 

\section{Exponential expansion}\label{EXPONENTIALEX}
For the exponential expansion, one has the scale factor given as \cite{ellis1991exact}
\begin{equation}\label{scalefactorexp}
a(t)=Ae^{wt}, \text{  where } A,w>0. 
\end{equation}
So we write Eq. \eqref{kineticterm} as
\begin{equation}\label{dotphi11}
\dot{\phi}^{2}=2K-\frac{4M}{3a^{4}}\; . 
\end{equation}
For $k=0$ and $k=-1$ we have a complex scalar field since both $a$ and $M$ are positives. For $k=1$ we can write Eq. \eqref{dotphi11} as
\begin{equation}
\dot{\phi}=\pm\left(\frac{2}{a^{2}}\right)^{1/2}\left(1-\frac{2M}{3a^{2}}\right)^{1/2}\; . 
\end{equation}
If $\frac{2M}{3a^{2}}>1$, we would have a complex term. We only consider the case where $\frac{2M}{3a^{2}}<<1$, this is because during
inflation epoch the radiation was too small to be considered during inflation, we therefore have
\begin{equation}\label{dotphik1}
\dot{\phi}=\pm\left(\frac{2}{a^{2}}\right)^{1/2}\left(1-\frac{M}{3a^{2}}\right)\; . 
\end{equation}
By using Eq. \eqref{scalefactorexp} in Eq. \eqref{dotphik1} and performing integration, we get
\begin{equation}\label{phiexponentialexample}
\phi(t)=\pm\left(-\frac{\sqrt{2}}{Aw}e^{-wt}+\frac{\sqrt{2}M}{9A^{3}w}e^{-3wt}-\phi_{0}\right)\;, 
\end{equation}
where $\phi_{0}$ is the constant of integration.
One can recover the scalar field solution obtained in \cite{ellis1991exact} by simply setting $M=0$ which means a radiation-less universe
consideration (de Sitter universe).  
\subsection{Reconstruction of $f(R)$ for exponential expansion models}
Combining Eqs. \eqref{Ricciscalar} and \eqref{scalefactorexp}, we write the Ricci scalar as 
\begin{equation}
R=12w^{2}+\frac{6}{A^{2}e^{2wt}}\; . 
\end{equation}
Thus we have time as function of $R$ as
\begin{equation}\label{tRexample1}
t(R)=\frac{1}{2w} \ln \left(\frac{6}{A^{2}(R-12w^{2})}\right)\; .
\end{equation}
In Eq.\eqref{tRexample1}, we should have $R>12w^{2}$ to avoid negative expression inside the braket.
Replacing Eq. \eqref{tRexample1} in Eq. \eqref{phiexponentialexample} we have 
\begin{equation}\label{phi(R)exp}
\phi(R)=\pm\Big[-\frac{\sqrt{2}}{Aw}\left(\frac{6}{A^{2}(R-12w^{2})}\right)^{-1/2} 
+\frac{\sqrt{2}M}{9A^{3}w}\left(\frac{6}{A^{2}(R-12w^{2})}\right)^{-3/2}-\phi_{0}\Big]\; . 
\end{equation}
Using the definitions in Eqs. \eqref{definitionofphi} and \eqref{defintionf(R)} in \eqref{phi(R)exp}, $f(R)$ can be written as
\begin{equation}
\begin{split}
f(R)&= \pm\Big[\frac{2}{3\sqrt{3}w}(R-12w^{2})^{3/2} +\frac{2\sqrt{2}M}{45\times 6^{3/2}w}\left(R-12w^{2}\right)^{5/2}-\phi_{0}R\Big]
+R+C_{1}\;,
\end{split}
\end{equation}
where $C_{1}$ is the constant of integration. For some limiting cases, this equation reduces to 
$f(R)=R$ which the Lagrangian for GR.
\subsection{Potential $V(\phi)$ for exponential expansion models}\label{POTENTIALEXPO} 
From Eq. \eqref{phiexponentialexample}, if we set
\begin{equation}\label{definitionofx}
x=e^{-wt} \;,
\end{equation}
then we write Eq. \eqref{phiexponentialexample} considering only the positive root with understanding that the
similar analysis can be done with the negative root as well
\begin{equation}
\phi+\phi_{0}=-\frac{\sqrt{2}k}{Aw}x+\frac{\sqrt{2}M}{9\sqrt{k}A^{3}w}x^{3}\; . 
\end{equation}
The above equation has two complex solutions and one real solution given as
\begin{equation}\label{solutionofx1}
\begin{split}
x&= 3^{\frac{2}{3}}Aw^{\frac{2}{3}}M^{\frac{1}{3}}\Big[3\sqrt{2}Aw(\phi+\phi_{0})+\sqrt{6}\big(3w^{2}(\phi^{2}+\phi^{2}_{0})
+6w^{2}\phi\phi_{0}+\frac{8A^{2}}{M}\big)^{1/2}\Big]^{1/3}  \\
&- 6^{\frac{2}{3}}A^{\frac{1}{3}}M^{-\frac{2}{3}}\Big[3\sqrt{2}Aw(\phi+\phi_{0})+\sqrt{6}\big(3w^{2}(\phi^{2}+\phi^{2}_{0})
+6w^{2}\phi\phi_{0}+\frac{8A^{2}}{M}\big)^{1/2}\Big]^{-1/3} \; .
\end{split}
\end{equation}
From Eq. \eqref{definitionofx}, we can write
\begin{equation}
t=-\frac{\ln x}{w}\; . 
\end{equation}
Here $\ln x$ is constrained to be negative such that we do not experience negative time parameter. This is because the 
parameter $w$ is a positive number by construction.
The potential defined in Eq. \eqref{potential} reads
\begin{equation}
V(t)=3w^{2}+ \frac{2}{A^{2}}e^{-wt}-\frac{M}{3A^{4}}e^{-4wt},
\end{equation}
so that we have $V(\phi)$ as
\begin{equation}
V(\phi)= 3w^{2}+ \frac{2}{A^{2}}x-\frac{M}{3A^{4}}x^{4}\;.
\end{equation}
This potential and its first and second derivatives with respect to the scalar field
$\phi$ will help in constructing the slow-roll parameters in Section \eqref{INFLATIONSEC}.

\section{Linear expansion}\label{LINEAREX}
For linear expansion, one has a scale factor given as \cite{ellis1991exact}
\begin{equation}\label{alinear}
a=At, 
\end{equation}
 where $A>0$. By replacing Eq. \eqref{alinear} in Eq. \eqref{kineticterm} we get
\begin{equation}
\dot{\phi}=\pm\left(\frac{2(k+A^{2})}{a^{2}}\right)^{1/2}\left(1-\frac{2M}{a^{2}(k+A^{2})}\right)^{1/2} \; .
\end{equation}
If $\frac{2M}{a^{2}(k+A^{2})}>1$, we would have a complex scalar field. We therefore consider the case where
$\frac{2M}{a^{2}(k+A^{2})}<<1$. This is assumed with the fact that $a$ is small and  the constant $A$ is big.  This leads to the approximation of the above equation as
\begin{equation}
\dot{\phi}=\pm\left[\left(\frac{2(k+A^{2})}{a^{2}}\right)^{1/2}-\frac{\sqrt{2}M}{a^{2}\sqrt{(k+A^{2})}}\right]\; . 
\end{equation}
Replacing back the expression for scale factor and performing integration, we have the scalar field given as
\begin{equation}\label{linearphit}
\phi(t)=\pm\Big[\frac{\sqrt{2(k+A^{2})}}{A}\ln(t) +\frac{\sqrt{2}M}{A^{2}\sqrt{k+A^{2}}t}-\phi_{0}\Big]\; . 
\end{equation}
\subsection{Reconstruction of $f(R)$ for linear inflation models}
In order to construct $f(R)$ models from scalar field formulations obtained in Eq. \eqref{linearphit}, we need to have $\phi(R)$. 
To achieve this we start from the definition of the Ricci scalar given in Eq.\eqref{Ricciscalar}, so that we can write 
\begin{equation}
 R=\frac{6(A^{2}+k)}{A^{2}t^{2}}\; .
\end{equation}
From this equation, one has $t(R)$ given as
\begin{equation}
t(R)=\left(\frac{6(A^{2}+k)}{A^{2}R}\right)^{1/2}\; . 
\end{equation}
Therefore, the scalar field in Eq. \eqref{linearphit} has the form
\begin{equation}\label{phi(R)linear}
\phi(R)=\pm\Big[\frac{\sqrt{2(A^{2}+k)}}{A} \ln\left(\frac{6(A^{2}+k)}{A^{2}R}\right)^{1/2} 
+\frac{\sqrt{2}M}{A^{2}\sqrt{A^{2}+k}}\left(\frac{6(A^{2}+k)}{A^{2}R}\right)^{-1/2}-\phi_{0}\Big]\; .
\end{equation}
Using the definitions in Eqs. \eqref{definitionofphi} and \eqref{defintionf(R)} in Eq. \eqref{phi(R)linear},  we have  $f(R)$ written as
\begin{equation}
\begin{split}
f(R)&=\pm\Big\{\frac{\sqrt{2(A^{2}+k)}}{2A(A^{2}+k)}\left[A^{2}R\ln\Big(\frac{6(A^{2}+k)}{A^{2}R}\Big)
+kR\ln\Big(\frac{6(A^{2}+k)}{A^{2}R}\Big) +A^{2}R+kR \right] \\
& +\frac{\sqrt{12}M R^{3/2}}{9A(A^{2}+k)}-\phi_{0}R\Big\}+R+C_{2}\; ,
\end{split}
\end{equation}
where $C_{2}$ is a constant of integration. This Lagrangian takes different form depending on the geometry of the 
spacetime.

\subsection{Potential $V(\phi)$ for linear inflation models}
The potential in terms of cosmic time $t$ is given as
\begin{equation}\label{linearVtt}
V(t)= \frac{2}{t^{2}}+\frac{2k}{A^{2}t^{2}}-\frac{M}{3A^{4}t^{4}}\; .
\end{equation}
We need to get $t(\phi)$ in order to get $V(\phi)$.
From Eq. \eqref{linearphit} with the consideration of its positive root, when one considers flat universe $(k=0)$, we have
\begin{equation}\label{lineart}
t=-\frac{M}{A^{3}\mathbf{W}(x)}\;, 
\end{equation}
where $x=-\frac{Me^{-\frac{\sqrt{2}(\phi+\phi_{0})}{2}}}{A^{3}}$ and $\mathbf{W}(x)$ is Lambert's function.
The Taylor expansion for Lambert's function is given as
\begin{equation}
\mathbf{W}(x)=\sum^{\infty}_{n=1}\frac{(-n)^{n-1}}{n!}x^{n}\approx x-x^{2}+\frac{3}{2}x^{3}-\frac{8}{3}x^{4}+ O(x^{5})\; . 
\end{equation}
Thus up to quadratic order, one has
\begin{equation}\label{Lambertkzero}
\mathbf{W}(x)= -\frac{Me^{-\frac{\sqrt{2}(\phi+\phi_{0})}{2}}}{A^{3}}-\frac{M^{2}e^{-\sqrt{2}(\phi+\phi_{0})}}{A^{6}}+... \; .
\end{equation}
Replacing Eq. \eqref{Lambertkzero} in  Eq. \eqref{lineart}, we have $t(\phi)$ as
\begin{equation}
t(\phi)=\frac{M}{A^{3}}\Big[\frac{Me^{-\frac{\sqrt{2}(\phi+\phi_{0})}{2}}}{A^{3}}
+\frac{M^{2}e^{-\sqrt{2}(\phi+\phi_{0})}}{A^{6}}\Big]^{-1} \;.
\end{equation}
Replacing $t(\phi)$ in Eq. \eqref{linearVtt}, we have
\begin{equation}
V(\phi)=2\Big[e^{-\frac{\sqrt{2}(\phi+\phi_{0})}{2}}
+\frac{Me^{-\sqrt{2}(\phi+\phi_{0)}}}{A^{3}}\Big]^{2}
-\frac{M}{3}\Big[\frac{e^{-\frac{\sqrt{2}(\phi+\phi_{0})}{2}}}{A^{2}}
+\frac{Me^{-\sqrt{2}(\phi+\phi_{0)}}}{A^{5}}\Big]^{4}\; . 
\end{equation}
For $k=\pm 1$, we have
\begin{equation}\label{vforkone}
V(t)=\frac{2(A^{2}\pm 1)}{A^{2}t^{2}}-\frac{M}{3A^{4}t^{4}}\; . 
\end{equation}
From Eq. \eqref{linearphit}, we have
\begin{equation}
t(\phi)=\exp\Big[\frac{A(\phi_{0}+\phi)}{\sqrt{2(A^{2}\pm 1)}}+\mathbf{W}(x)\Big], 
\end{equation}
where $x=-\frac{Me^{-\frac{A(\phi+\phi_{0})}{\sqrt{2(A^{2}\pm 1)}}}}{A(A^{2}\pm 1)}$ and the new 
$\mathbf{W}(x)$ is expanded as 
\begin{equation}
\mathbf{W}(x)= -\frac{M}{A(A^{2}\pm 1)}e^{-\frac{A(\phi+\phi_{0})}{\sqrt{2(A^{2}\pm 1)}}}
-\frac{M^{2}}{A^{2}(A^{2}\pm 1)^{2}}e^{-2\frac{A(\phi+\phi_{0})}{\sqrt{2(A^{2}\pm 1)}}}+...\; .
\end{equation}
Thus, one has $t(\phi)$ given as
\begin{equation}
t(\phi)=\exp\Big[\frac{A(\phi_{0}+\phi)}{\sqrt{2(A^{2}\pm 1)}}-\frac{M}{A(A^{2}\pm 1)}e^{-\frac{A(\phi+\phi_{0})}{\sqrt{2(A^{2}\pm 1)}}}
-\frac{M^{2}}{A^{2}(A^{2}\pm 1)^{2}}e^{-2\frac{A(\phi+\phi_{0})}{\sqrt{2(A^{2}\pm 1)}}}\Big]\; . 
\end{equation}
Replacing this expression for $t(\phi)$ in Eq. \eqref{vforkone}, we have
\begin{equation}
\begin{split}
 V(\phi)&= \frac{2(A^{2}\pm 1)}{A^{2}}\exp\Big\{-2\Big[\frac{A(\phi_{0}+\phi)}{\sqrt{2(A^{2}\pm 1)}}
 -\frac{M}{A(A^{2}\pm 1)}e^{-\frac{A(\phi+\phi_{0})}{\sqrt{2(A^{2}\pm 1)}}}
-\frac{M^{2}}{A^{2}(A^{2}\pm 1)^{2}}e^{-2\frac{A(\phi+\phi_{0})}{\sqrt{2(A^{2}\pm 1)}}}\Big]\Big\}\\
&-\frac{M}{3A^{4}}\exp\Big\{-4\Big[\frac{A(\phi_{0}+\phi)}{\sqrt{2(A^{2}\pm 1)}}-\frac{M}{A(A^{2}\pm 1)}e^{-\frac{A(\phi+\phi_{0})}{\sqrt{2(A^{2}\pm 1)}}}
-\frac{M^{2}}{A^{2}(A^{2}\pm 1)^{2}}e^{-2\frac{A(\phi+\phi_{0})}{\sqrt{2(A^{2}\pm 1)}}}\Big]\Big\}\; .
\end{split}
\end{equation}
In the following section, we apply the results obtained so far to the inflation epoch especially on slow-roll approximation to see
how the models respond to inflation parameters $n_{s}$ and $r$.
\section{Application to inflation epoch}\label{INFLATIONSEC}

\begin{figure}[h!]
  \centering
    \subfloat[Plot of $r(w)$ for $M=1$, $A=15$, $\phi_{0}= 1$ and 
    $\phi=1.2$, for exponential expansion, the red line refers to the Planck data.]
 {\includegraphics[width=0.45\textwidth]{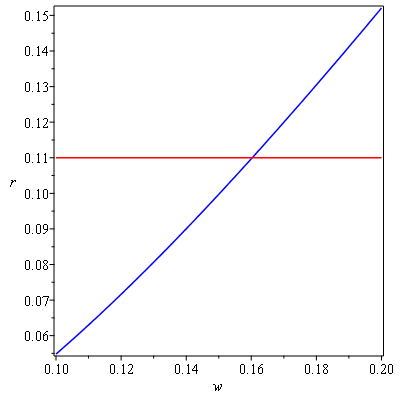}\label{figrw}}
  \hfill
  \subfloat[Plot of $n_{s}(w)$ for $M=1$, $A=15$, $\phi_{0}= 1$ and $\phi=1.2$, for exponential expansion, 
  the red and green lines  refer to the upper and lower bounds of Planck data respectively.]{\includegraphics[width=0.45\textwidth]{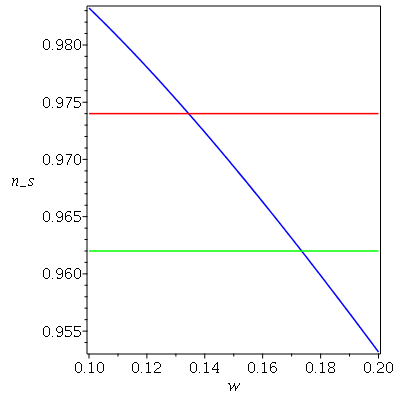}\label{fignsw}}
  \caption{}
\end{figure}

\begin{figure}[h!]
  \centering
    \subfloat[Plot of $r(A)$ for $M=1$, $w=0.15$, $\phi_{0}= 1$ and 
    $\phi=1.5$, for exponential expansion, the red line  refers to the Planck data.]
 {\includegraphics[width=0.45\textwidth]{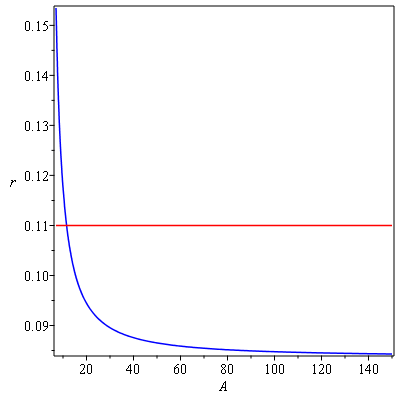}\label{figrA}}
  \hfill
  \subfloat[Plot of $n_{s}(A)$ for $M=1$, $w=0.15$, $\phi_{0}= 1$ and 
    $\phi=1.5$, for exponential expansion, the red and green lines  refer to the upper
    and lower bounds of Planck data respectively.]{\includegraphics[width=0.45\textwidth]{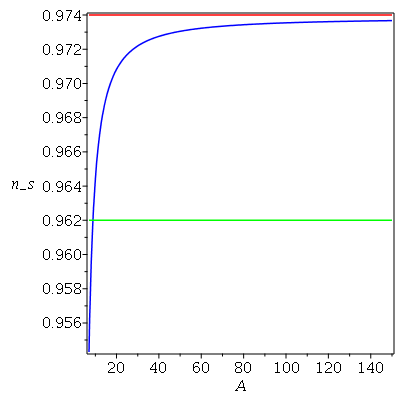}\label{fignsA}}
  \caption{}
\end{figure}

\begin{figure}[h!]
  \centering
    \subfloat[Plot of $r(\phi_{0})$ for $M=1$, $w=0.15$, $A= 15$ and 
    $\phi=1.15$, for exponential expansion, the red line refers to the Planck data.]
 {\includegraphics[width=0.45\textwidth]{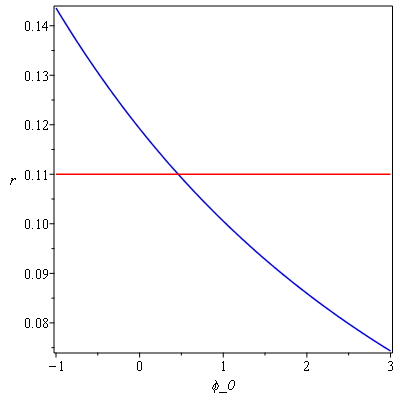}\label{figrphi0}}
  \hfill
  \subfloat[Plot of $n_{s}(\phi_{0})$ for $M=1$, $w=0.15$, $A= 15$ and 
    $\phi=1.15$, for exponential expansion, the red and green lines  refer to the upper and lower bounds of Planck data respectively.]{\includegraphics[width=0.45\textwidth]{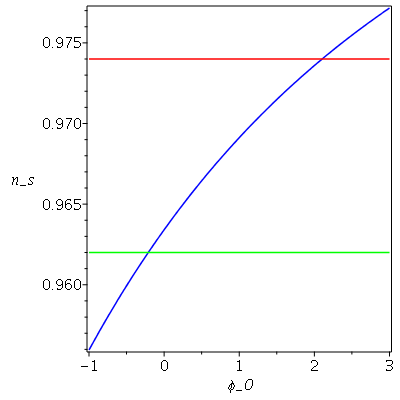}\label{fignsphi0}}
  \caption{}
\end{figure}

\begin{figure}[h!]
  \centering
    \subfloat[Plot of $r(\phi)$ for $M=1$, $w=0.15$, $A= 15$ and 
    $\phi_{0}=1.15$, for exponential expansion, the red line refers to the Planck data.]
 {\includegraphics[width=0.45\textwidth]{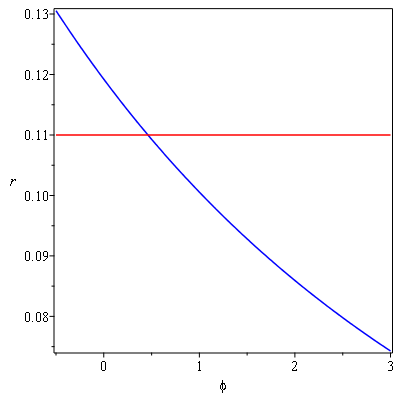}\label{figrphi}}
 \hfill
  \subfloat[Plot of $n_{s}(\phi)$ for $M=1$, $w=0.15$, $A= 15$ and 
    $\phi_{0}=1.15$, for exponential expansion, the red and green lines  refer to the upper and lower bounds of Planck data respectively.]{\includegraphics[width=0.45\textwidth]{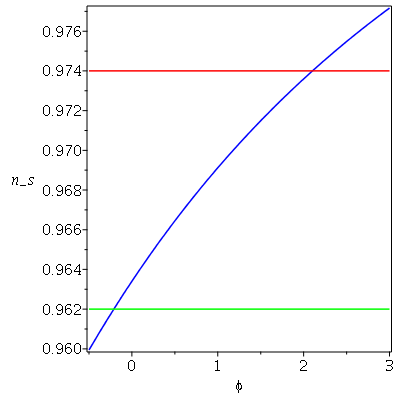}\label{fignsphi}}
  \caption{}
\end{figure}
\begin{figure}[h!]
  \centering
    \subfloat[Plot of $r(A)$ for $M=1$, $\phi_{0}= 750$ and 
    $\phi=1.05$, for linear expansion, the red line refers to the Planck data.]
 {\includegraphics[width=0.45\textwidth]{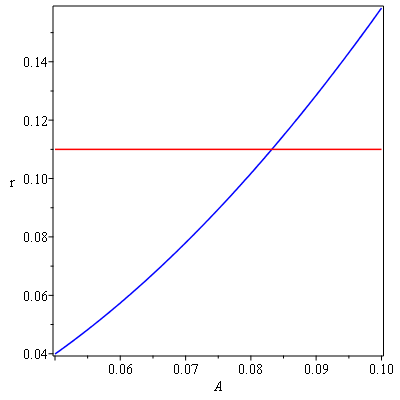}\label{figrlinear}}
  \hfill
  \subfloat[Plot of $n_{s}(A)$ for $M=1$, $\phi_{0}= 750$ and 
    $\phi=1.05$, for linear expansion , the red and green lines  refer to the upper and lower bounds of Planck data respectively.]{\includegraphics[width=0.45\textwidth]{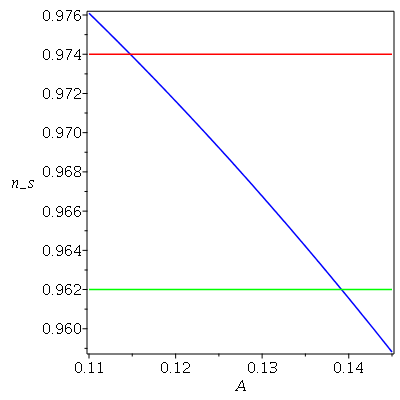}\label{fignslinear}}
  \caption{}
\end{figure}

We focus our interests to slow-roll approximations. The slow-roll approximation has
 the condition that the kinetic energy of the scalar field is much less than the potential \cite{liddle1994formalizing}. 
 This leads to two conditions as
\ber
&&\dot{\phi}^{2}<V(\phi)\; , \label{slow1}\\
&&2|\ddot{\phi}|<|V'(\phi)|\; ,\label{slow2}
\eer
where here $'$ indicates differentiation with respect to the scalar field $\phi$.
Then we can define two  potential slow-roll parameters $\epsilon (\phi)$ and $\eta(\phi)$ \cite{liddle1994formalizing,liddle1992cobe}
\begin{equation}
\epsilon(\phi)=\frac{1}{2\kappa^{2}}\Big(\frac{V'(\phi)}{V(\phi)}\Big)^{2},
\end{equation}
and
\begin{equation}
\eta(\phi)=\frac{1}{\kappa^{2}}\Big(\frac{V''(\phi)}{V(\phi)}\Big)\;.
\end{equation}
The spectral index $n_{s}$ and the tensor-to-scalar ratio $r$ are defined, respectively,  as \cite{liddle1994formalizing,liddle1992cobe}
\ber
&&n_{s}=1-6\epsilon+2\eta,\label{ns}\;,\\
&&r=16\epsilon\label{r}\;.
\eer

The numerical computation of these two parameters is done for both exponential and linear expansion models. 
The observational values of $n_{s}$ and $r$ are taken from the Planck data \cite{ade2016planck}.

The following figures summarize the change of the parameters $w,A,\phi_{0}$ and $\phi$ to get
both spectral index $n_{s}$ and tensor-to-scalar ratio $r$ close to the observational values
for the linear law of expansion for the case when $k=1$. 
Where $k=0$ and $k=-1$, one has negative and complex values for the spectral index $n_{s}$ and the tensor-to-scalar ratio $r$, this evidently
makes the corresponding $f(R)$ Lagrangian being ruled out from the viable ones.

It is clear from the figures presented in this work that
the observational results of $n_{s}$ and $r$ constrain the ranges of the parameters under variation. 
In Figure \eqref{figrw} one can notice that $r$ increases monotonically as function of $w$ until it surpasses the upper limit of Planck data which 
constrain that $r<0.11$. On the other side in Figure \eqref{fignsw} the spectral index decreases as $w$ changes and the only  acceptable
values can be obtained between $w=0.14$ and $w=0.177$.   
In Figure \eqref{figrA}, after $n_{s}$ crosses the observational value, it asymptotically varies with increasing of
$A$. On the other side, in Figure \eqref{fignsA}, $n_{s}$ increases and saturated to a finite amplitude  after it crosses
the lower bound from the Planck data.
In Figure \eqref{figrphi0}, the tensor-to-scalar ratio $r$ is decreasing with increasing of $\phi_{0}$ and the values that are
compatible with observation are achieved after passing $\phi_{0}=0.42$.
In Figure \eqref{fignsphi0}, the spectral index $n_{s}$ increases with increase of $\phi_{0}$ until it crosses the lower and the 
upper boundaries of the observational data. The same behavior is manifested in Figure \eqref{figrphi} and Figure \eqref{fignsphi} 
for $r$ and $n_{s}$ for changing $\phi$.

For linear expansion model, the Figures \eqref{figrlinear} and \eqref{fignslinear} show how $r$
and $n_{s}$ change as function of $A$. The tensor-to-scalar ratio increases as function of $A$ and $n_{s}$ decreases  as function of $A$.
This makes $n_{s}$ cross the upper bound first and later it crosses the lower bound.

\section{Conclusions }\label{CONCLUSION}

In this paper, we studied cosmological fluid systems composed of radiation and a scalar field and with predetermined inflationary expansion scenarios of the FLRW background spacetime.  Using the scalar field as the extra degree of freedom that appears in generic $f(R)$-gravity models, we reconstructed  $f(R)$ Lagrangians and the corresponding potentials for the scalar field. The reconstructed $f(R)$  actions have GR and $\Lambda$CDM actions as limiting cases.
We applied the results to slow-roll approximations to show that actually, the parameters that are involved in the Lagrangian reconstruction can be constrained by observational data.

\section*{Acknowledgements}
This work is based on the research supported in part by the National Research Foundation of South Africa (Grant Number 109257).
 JN gratefully acknowledges financial support from the Swedish International Development Cooperation Agency (SIDA) 
through the International Science Program (ISP) to the University of Rwanda (Rwanda Astrophysics, Space and Climate Science Research Group),
and Physics Department, North-West University, Mafikeng
Campus, South Africa, for hosting him during the preparation of this paper. HS acknowledges the partial financial support from African Institute for Mathematical Sciences (AIMS-Ghana).

\end{document}